\documentclass[twocolumn,english,prbl,twocolumn,floatfix,footinbib,superscriptaddress,showpacs,showkeys,notitlepage]{revtex4-1}
\usepackage[T1]{fontenc}
\usepackage[latin9]{inputenc}
\usepackage{babel}
\usepackage{amsmath}
\usepackage{graphicx}
\usepackage{esint}
\usepackage{hyperref}
\usepackage{babel}
\usepackage{xcolor}
\usepackage{bibentry}

\definecolor{applegreen}{rgb}{0.55, 0.71, 0.0}
\newcommand{\sk}[1]{{#1}}

\global\long\def\ket#1{\left| #1\right\rangle }
\global\long\def\bra#1{\left\langle #1 \right|}

\global\long\def\av#1{\left\langle #1 \right\rangle }
\global\long\def\tr{\text{tr}}

\global\long\def\pd{\partial}

\global\long\def\bs#1{\boldsymbol{#1}}
\global\long\def\t#1{\text{#1}}

\begin{document}

\title{Spin transfer torque induced paramagnetic resonance }

\author{Alexey M. Shakirov}
\email[]{a.shakirov@rqc.ru}

\selectlanguage{english}%

\affiliation{Russian Quantum Center, Novaya Street 100A, 143025 Skolkovo, Moscow
Region, Russia}

\affiliation{Department of Physics, Lomonosov Moscow State University, Leninskie
Gory 1, 119992 Moscow, Russia}

\author{Alexey N. Rubtsov}

\affiliation{Russian Quantum Center, Novaya Street 100A, 143025 Skolkovo, Moscow
Region, Russia}

\affiliation{Department of Physics, Lomonosov Moscow State University, Leninskie
Gory 1, 119992 Moscow, Russia}

\author{Pedro Ribeiro}

\affiliation{CeFEMA, Instituto Superior T\'ecnico, Universidade de Lisboa, Av. Rovisco
Pais, 1049-001 Lisboa, Portugal}

\affiliation{Beijing Computational Science Research Center, 100193 Beijing, China}
\begin{abstract}
We show how the spin-transfer torque generated by an ac voltage may
be used to excite a paramagnetic resonance of an atomic spin deposited
on a metallic surface. This mechanism is independent of the environment
of the atom and may explain the ubiquity of the paramagnetic resonance
reported by Baumann \textit{et al.} [\href{http://dx.doi.org/10.1126/science.aac8703}{Science \textbf{350}, 417 (2015)}]. The current and spin dynamics
are modeled by a time-dependent Redfield master equation generalized
to account for the periodic driven voltage. Our approach shows that
the resonance effect is a consequence of the nonlinearity of the
coupling between the magnetic moment and the spin-polarized current
which generates a large second-harmonic amplitude that can be measured
in the current signal. 
\end{abstract}
\maketitle

\section{Introduction \label{sec:Introduction}}

The interplay between the electronic motion and spin degrees of freedom
is a key ingredient to design atomic-size magnetic devices. Recently, 
experimental advances have allowed a number of remarkable results
in nonmagnetic control over single magnetic molecules and
other artificially fabricated spin structures \cite{Editorial2015,Zhang2015}. This progress
is significantly driven by future-technology demands, as single magnetic
atoms have long been viewed as structural elements for high-density
information storage and processing devices \cite{Troiani2005,Imre2006,Bogani2008,Khajetoorians2011,Natterer2017}.
Recently, these structures were shown to preserve quantum coherence
under certain circumstances \cite{Heinrich2013,Lidar2014}, which triggered
a renewed interest in their use for quantum information processing
\cite{Ghirri2017}.

Individual addressing of the atomic spin can be achieved only by non-magnetic
means, using an external current applied locally at the atomic site \cite{Meier2008,Wiesendanger2009,Loth2010}.
The degree of control over the atomic spin is ultimately determined
by the nature of the coupling between the current and the magnetic
moment. Therefore, it is of foremost importance to explore different
coupling mechanisms. Reference \cite{Komeda2011} reports the current
control of TbPc$_{2}$ magnetic properties by applying controlled
current pulses via a scanning tunneling microscope (STM) setup. The physical origin of this effect is a current-induced molecular conformation. Recently, Baumann \textit{et
al.} \cite{Baumann2015} have reported the induction of a paramagnetic resonance of individual
magnetic atoms on a surface by application of an alternating current. 
Several proposals have been put forward to understand
this effect \cite{Baumann2015,Berggren2016,Lado2017}, which involves
coupling the individual magnetic atom to mechanical (orbital) degrees
of freedom that in turn couple to the alternating field. However, such coupling
should be highly dependent on the local environment of the atom and
thus these scenarios have difficulties explaining the ubiquity of the
observations. 

In this paper, we show that a current-induced spin torque can effectively couple the spin of the magnetic atom to a locally applied alternating voltage.
By modeling the experimental setup of Ref. \cite{Baumann2015},
we argue that this mechanism can be responsible for the observed paramagnetic
resonance and is compatible with the measured decoherence and decay
times. 
The dissipative nature of the current-induced spin torque renders this effect distinct from mechanisms that induce a time dependence to the Hamiltonian of the local moment.

\begin{figure}[tb]
\includegraphics[width=1\columnwidth]{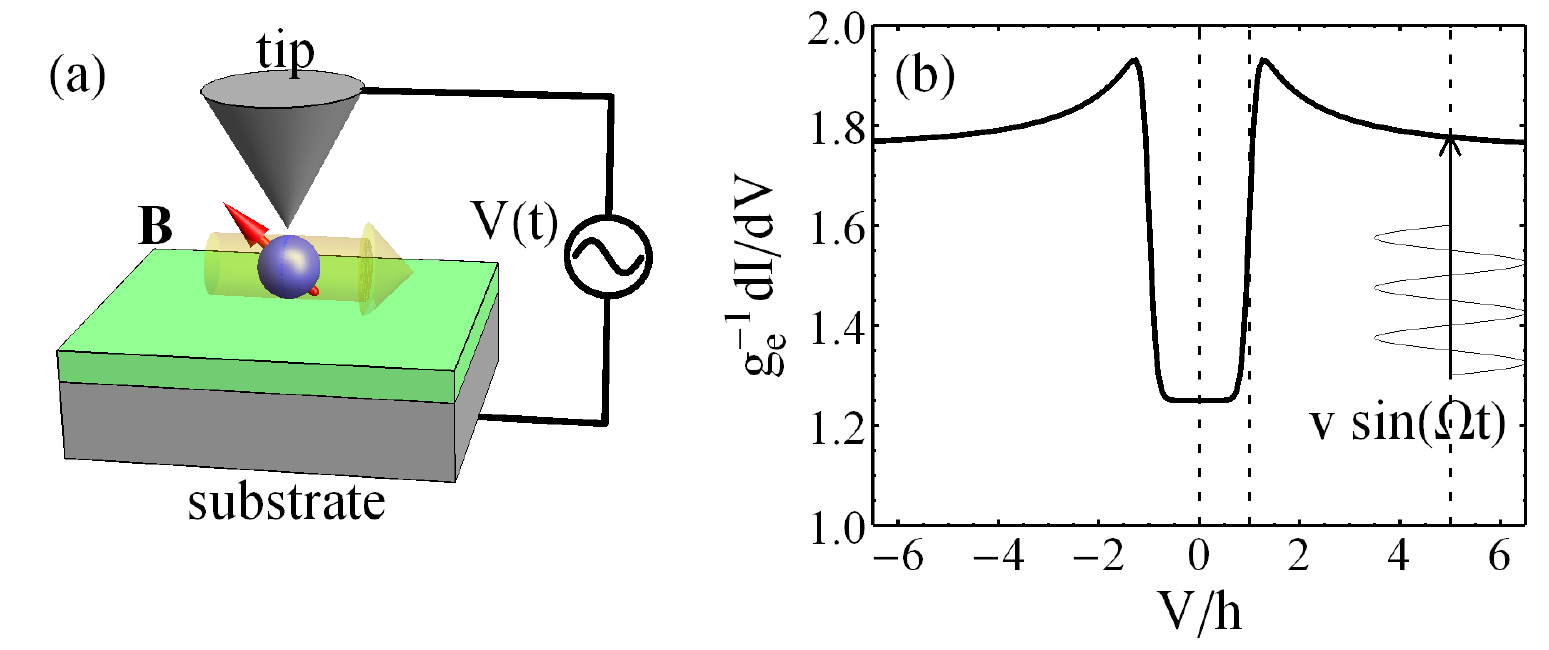}\caption{(a) The sketch of the setup that consists of the magnetic atom, the
metallic substrate coated by the insulating layer, and the polarized
tip. The external magnetic field $\boldsymbol{B}$ acts on the atom,
and the periodically changing voltage $V\left(t\right)=V+v\sin\left(\Omega t\right)$
is applied across the tip. (b) The dc conductance spectrum of the
setup. The sine curve indicates the range of values of the instant
voltage.}

\label{fgr:Setup} 
\end{figure}

Baumann \textit{et al.} \cite{Baumann2015} deposited a single magnetic atom
(Fe or Co) on a metallic surface, Ag(001), coated with a thin insulator,
an atomically thick MgO layer. A sketch of the setup is given in Fig.
\ref{fgr:Setup}. With a STM tip placed on top of the atom, an alternating
voltage ($\sim$ 2 -- 3 MHz) was applied between the tip and the metallic
substrate, in addition to a direct voltage component. While swiping
the frequency of the applied ac signal, a peak in the dc current response
was observed marking a single magnetic excitation. The effect was
shown to be present for a spin-polarized tip on Fe atoms and absent
for Co atoms or for a spin-unpolarized tip. The width of the resonance
measured in the current signal was related with $T_{2}$ and shown
to be much smaller then the measured relaxation time $T_{1}$. Similar results were subsequently reported in Refs. \cite{Yang2017} and \cite{Willke2018}.

Although we do not intend to reproduce the exact conditions of the experiment, we 
consider below a minimal model able to capture the main physical effects. 

\section{Model \& Method \label{sec:Model}}

A description of magnetic atoms and molecules in terms of an effective
spin Hamiltonian emerges in the presence of large charge gap that
effectively promotes the number of atomic electrons to a good quantum
number. At low temperatures, in the regime of weak hybridization between
the localized orbitals and the nearby itinerant electrons, tunneling
arises by virtual excitations of the localized charge state giving
rise to an effective exchange interaction between the localized spin
and that of the electronic environment. This situation can be modeled
by the Hamiltonian $H=H_{S}+H_{E}+H_{C}$, which includes the atomic
subsystem $H_{S}$, the electronic degrees of freedom of the environment
$H_{E}$, and the coupling term $H_{C}$. Here, the electronic environment
consists of two metallic leads (substrate and tip) in thermal equilibrium
with inverse temperature $\beta$ and chemical potentials $\mu_{s}=0$
and $\mu_{t}=-V$, where $V$ is the dc component of the applied voltage,
and we set $e=1$ for the electron charge $-e$. The full voltage
across the device 
\begin{align}
V\left(t\right) & =V+v\sin\left(\Omega t\right)\label{eq:VoltageDriving}
\end{align}
also includes an alternating component with amplitude $v$ and frequency
$\Omega$. We account for this component by adding a constant shift
$-\Delta_{\nu}\sin\left(\Omega t\right)$, with $\Delta_{t}=v$ and
$\Delta_{s}=0$, to the energies of the lead states, which renders
the Hamiltonian of the tip time-dependent. The leads are also characterized
by spin-polarized density of states (DOS) that account for the tip
polarization and a bandwidth $W$, much larger than any energy scale
of the system (for further details see Ref. \cite{Shakirov2016a}).
The degree of the tip polarization is determined by the parameter
$p$ that ranges from $-1$ to $1$. The atomic system consists of
a single atom with a well-defined total spin $S=1/2$ in the presence
of an external magnetic field $\bs B$,
\begin{equation}
H_{S}=\boldsymbol{h}\cdot\boldsymbol{S},
\end{equation}
where $\boldsymbol{h}=g\mu_{B}\boldsymbol{B}$ is proportional to
the atomic $g$-factor and the Bohr magneton $\mu_{B}$. The system-environment
coupling Hamiltonian is given by the exchange interaction terms \cite{Kim2004,Fernandez-Rossier2009}
\begin{align}
H_{C} & =\sum_{a\nu\nu'}\sqrt{J_{\nu}^{a}J_{\nu'}^{a}}S_{a}\otimes s_{a}^{\nu\nu'},\label{eq:CouplingHamiltonian}\\
s_{a}^{\nu\nu'} & =\frac{1}{\sqrt{\mathcal{N}_{\nu}\mathcal{N}_{\nu'}}}\sum_{kk'ss'}c_{\nu ks}^{\dagger}\frac{\sigma_{ss'}^{a}}{2}c_{\nu'k's'},
\end{align}
where $a=0,x,y,z$, the axis $z$ is aligned with the tip polarization,
and $\sigma^{a}$ are the Pauli matrices (with $\sigma^{0}=S_{0}=I$).
Therefore, terms with $a=0$ correspond to the elastic tunneling of
electrons between the leads. The inelastic coupling is isotropic,
i.e., $J_{\nu}^{a}=J_{\nu}$. In the following, we use dimensionless coupling parameters $\varGamma^{a}_{\nu}=\pi J^{a}_{\nu}/2W$\sk{, with $W$ the bandwidth of the reservoirs, and consider only the isotropic case  $\varGamma^{a}_{\nu}=\varGamma_{\nu}$. }

To capture spin-torque effects, we employ a master-equation description
for the evolution of the reduced density matrix of the local moment that crucially includes the coherences. Therefore, we generalized
the Redfield master equation approach, previously used to model coherent
evolution and transport in engineered atomic spin devices \cite{Shakirov2016a,Shakirov2017},
to deal with the ac driving bias.

Following a standard procedure \cite{Breuer2002}, a master equation
$\pd_{t}\rho=\mathcal{L}_{t}\rho$ can be derived for the density
matrix of the atomic system, where $\mathcal{L}_{t}$ is a superoperator
of the Redfield type given by 
\begin{align}
\mathcal{L}_{t}\rho= & -i\left[H'_{S},\rho\right]-\sum_{aa'}\left[S_{a},\varLambda_{aa'}\left(t\right)\rho-\rho\varLambda_{aa'}^{\dagger}\left(t\right)\right].\label{eq:RedfieldSuperoperator}
\end{align}
Here $H'_{S}=H_{S}+\frac{1}{2}pJ_{t}S_{z}$ is the renormalized Hamiltonian
of the system, and 
\begin{align}
\varLambda_{aa'}\left(t\right)= & \sum_{\nu\nu'\alpha\alpha'}u_{aa'}^{\nu\nu'}\kappa_{\nu\nu'}^{t}\left(\omega_{\alpha}-\omega_{\alpha'}\right)\label{eq:LambdaOperators}\\
 & \times\ket{\alpha}\bra{\alpha}S_{a'}\ket{\alpha'}\bra{\alpha'},\nonumber 
\end{align}
where
\sk{$u_{aa'}^{\nu\nu'}  =\frac{\sqrt{\varGamma_{\nu}^{a}\varGamma_{\nu'}^{a}\varGamma_{\nu}^{a'}\varGamma_{\nu'}^{a'}}}{4\pi}\tr\left[\left(1+p_{\nu}\sigma^{z}\right)\sigma^{a}\left(1+p_{\nu'}\sigma^{z}\right)\sigma^{a'}\right]$}, and $\ket{\alpha}$ are eigenstates of $H'_{S}$ with energies
$\omega_{\alpha}$. The time dependence enters $\mathcal{L}_{t}$
via the quantity
\begin{align}
\kappa_{\nu\nu'}^{t}\left(\omega\right)= & \sum_{m=-\infty}^{\infty}i^{m}e^{im\Omega t}e^{i\frac{\Delta_{\nu}-\Delta_{\nu'}}{\Omega}\cos(\Omega t)}\label{eq:KappaFunctionTime}\\
 & \times J_{m}\left(-\frac{\Delta_{\nu}-\Delta_{\nu'}}{\Omega}\right)\kappa\left(\omega+m\Omega-\mu_{\nu}+\mu_{\nu'}\right),\nonumber 
\end{align}
where $J_{m}\left(x\right)$ are Bessel functions, that is the generalization
of the one obtained in the time independent case \cite{Shakirov2016a}
\begin{align}
\kappa\left(\omega\right) & =\frac{g\left(\beta\omega\right)+i\,f\left(\beta\omega\right)}{\beta}-\frac{i}{\pi}\omega\ln\frac{\left|\omega\right|}{cW},\label{eq:KappaFunction}
\end{align}
where $c$ is a constant of order $1$, $g\left(x\right)=x/(e^{x}-1)$,
and $f\left(x\right)=\frac{1}{\pi}P\int dy\,\left[g\left(y\right)+y\Theta\left(-y\right)\right]/\left(x-y\right)$.
The details of the derivation are given in Appendix \ref{sec:Appendix} and are a generalization of the method of Ref. \cite{Shakirov2016a}, obtained for a static
voltage (i.e., $v=0$), when $\kappa_{\nu\nu'}^{t}\left(\omega\right)=\kappa\left(\omega-\mu_{\nu}+\mu_{\nu'}\right)$
and the operators $\varLambda_{aa'}\left(t\right)$ are time independent.
For simplicity, the calculations below do not take into account the imaginary part of $\kappa\left(\omega\right)$. 
It is worth noting that this term may induce unphysical dynamics of the density matrix for large and moderate system-environment coupling, while it does not qualitatively change observables for weak coupling \cite{Shakirov2016a}.

The average value of the current between the leads can be obtained
introducing a counting field in the master equation (see Ref. \cite{Nazarov2003} and Appendix \ref{sec:Appendix}), or using a charge-specific formalism \cite{Rammer2004,Flindt2005,Shakirov2016a}
adapted to the time-dependent case, and is given by
\begin{align}
I\left(t\right) & =-\tr\sum_{aa'}\left[J_{aa'}\left(t\right)\rho\left(t\right)S_{a}+S_{a}\rho\left(t\right)J_{aa'}^{\dagger}\left(t\right)\right],\label{eq:AverageCurrent}
\end{align}
where the operators $J_{aa'}\left(t\right)$ are defined as 
\begin{align}
J_{aa'}\left(t\right) & =\sum_{\nu\nu'\alpha\alpha'}u_{aa'}^{\nu\nu'}\kappa_{\nu\nu'}^{t}\left(\omega_{\alpha}-\omega_{\alpha'}\right)\\
 & \times\left(\delta_{\nu t}-\delta_{\nu't}\right)\ket{\alpha}\bra{\alpha}S_{a'}\ket{\alpha'}\bra{\alpha'}.\nonumber 
\end{align}
The expression Eq. (\ref{eq:AverageCurrent}) has the same form as the
one obtained for the static case in Ref. \cite{Shakirov2017}, except
for the explicit time dependence of the density matrix and the operators
$J_{aa'}\left(t\right)$ due to the driving. Note that the current
obtained in this way assumes that the ac voltage has been turned on
in the infinite past and that the system has already attained a
periodic regime with the frequency of the drive. In practice, this
means that the duration of the ac pulse is considered to be larger
than the characteristic relaxation times. The average value of the
current in Eq. (\ref{eq:AverageCurrent}) can be separated into three
components \cite{Fernandez-Rossier2009,Delgado2010a}: (i) the elastic
component $I^{\left(1\right)}$ arising from the terms with $a=a'=0$,
(ii) the magnetoresistive component $I^{\left(2\right)}$ arising
from the terms with $a=0$ and $a'\neq0$, or $a\neq0$ and $a'=0$,
and (iii) the inelastic component $I^{\left(3\right)}$ arising from
the terms with $a\neq0$ and $a'\neq0$. We note that the elastic
component has the trivial dependence on the voltage $I^{\left(1\right)}\left(t\right)=g_{e}V\left(t\right)$,
satisfying the Ohm's law with $g_{e}=\varGamma_{s}^{0}\varGamma_{t}^{0}/\pi$.

\section{Results \label{sec:Results}}

We now apply the developed theory to study the electronic paramagnetic
resonance in magnetic atoms. We first calculate the ac spectra of
the current to demonstrate the appearance of the resonance peaks observed
in Ref. \cite{Baumann2015}; furthermore, we investigate how the spin dynamics
behaves in the vicinity of the resonance.

\begin{figure}[tb]
\includegraphics[width=1\columnwidth]{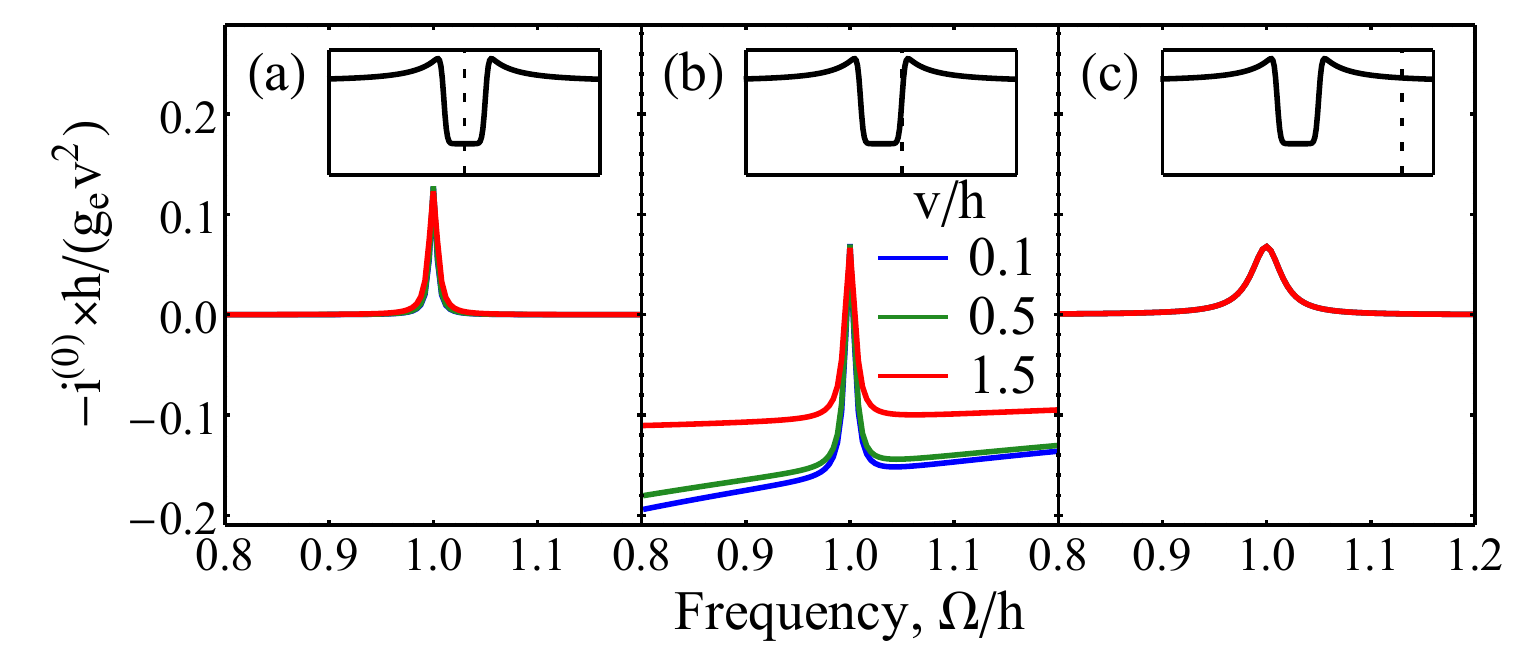}\caption{Dependence of $i^{\left(0\right)}$ term in the ac component of the
current Eq. (\ref{eq:CurrentHarmonics}) on the driving frequency for
different values of the dc voltage and the driving amplitude. We use
$h=0.1$ meV, $T=0.05h$, $\varGamma^{a}_{\nu}=0.1$, and (a) $V=0$, (b) $V=h$, (c) $V=5h$.
Since $i^{\left(0\right)}$ scales as $v^{2}$, we normalize it appropriately.
The insets show the spectrum of the dc current and the values of $V$
around which the voltage is driven.}
\label{fgr:Current} 
\end{figure}

For clarity, we split the average value of the current as 
\begin{align}
I\left(t\right) & =I+i\left(t\right),
\end{align}
where $I$ is the stationary current at constant voltage $V$, and
$i\left(t\right)$ is the differential response to the ac component.
Since the applied voltage Eq. (\ref{eq:VoltageDriving}) changes periodically,
$i\left(t\right)$ admits the Fourier series decomposition:
\begin{align}
i\left(t\right) & =i^{\left(0\right)}+\sum_{m=1}^{\infty}i^{\left(m\right)}\sin\left(m\Omega t+\phi_{m}\right).\label{eq:CurrentHarmonics}
\end{align}
Following Baumann \textit{et al.} \cite{Baumann2015}, we first study $i^{\left(0\right)}$,
whose dependence on the driving frequency for different values of
the dc voltage $V$ and the driving amplitudes $v$ is shown in Fig.
\ref{fgr:Current}. When the spin polarization of the current is perpendicular
to the magnetic field applied to the atomic spin, we see that in all
cases there is a pronounced peak at the resonant driving frequency $\Omega=h$.
As in the experiment, such peak is not observed if the current polarization
is collinear with the magnetic field $\boldsymbol{h}$. Note that
$i^{\left(0\right)}$ in Fig. \ref{fgr:Current} is normalized by
$v^{2}$, therefore, the collapse of these curves near the resonance
frequency, for different driving amplitudes $v$ and different values
of $V$, indicates that the non-linear processes generating the $i^{\left(0\right)}$
response are predominantly of second order in $v$. Away from the
resonance the $i^{\left(0\right)}$ response drops sharply for $V=0$
and $5h$. However, for $V=h$ one observes a non-zero $i^{\left(0\right)}$
response even off-resonance, this arises since the driving is done
around the dc voltage that corresponds to a highly non-linear part
of the spectrum as can be seen in the inset of Fig. \ref{fgr:Current}(b).

\begin{figure}[tb]
\includegraphics[width=\columnwidth]{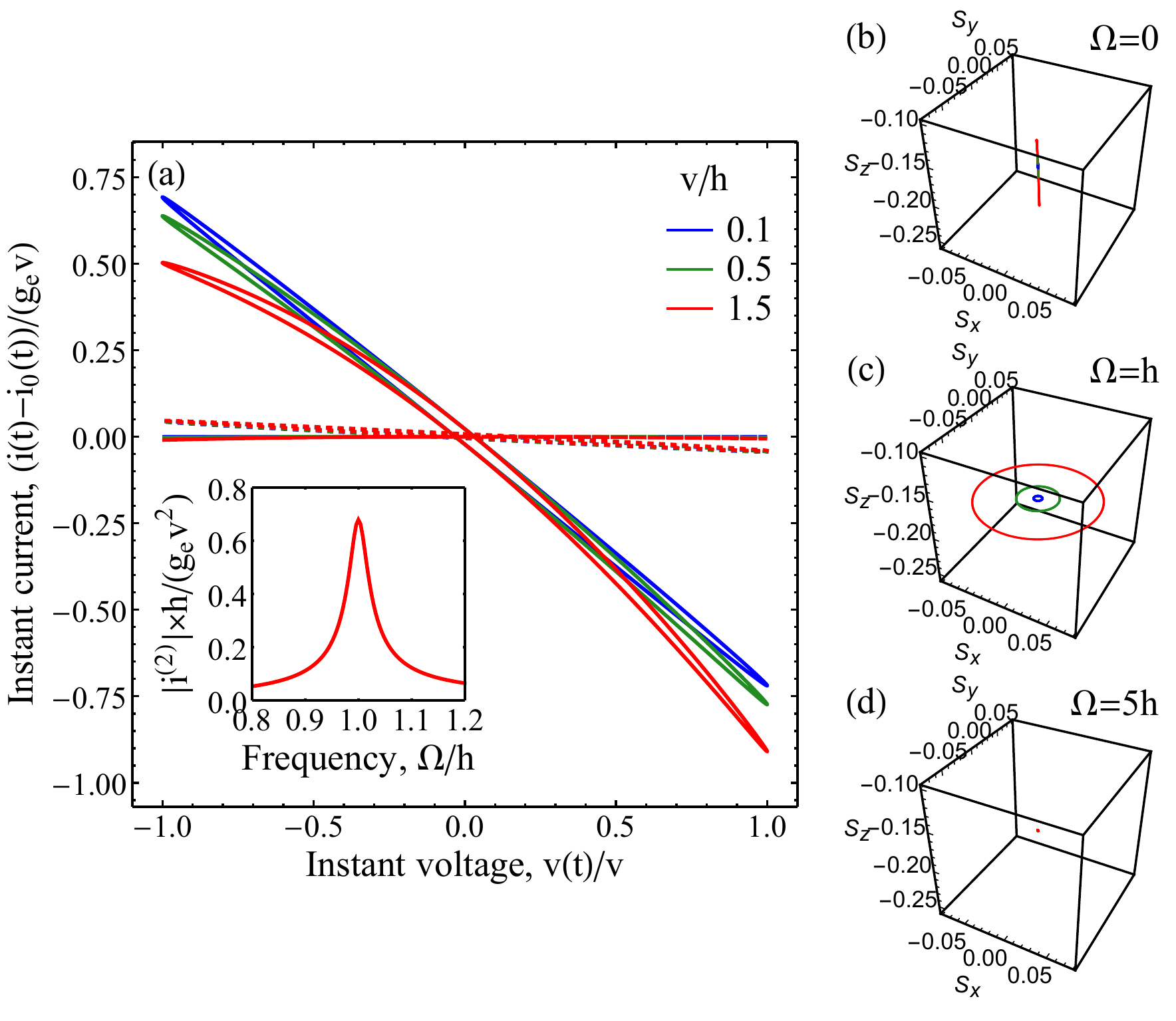}\caption{(a) Lissajous curves for $i\left(t\right)-i_{0}\left(t\right)$ versus $v\left(t\right)=v\,\sin\Omega t$ and (b)--(d) over-period trajectories of the average atomic spin at different amplitudes and frequencies of the driving. Dashed, solid, and dotted curves correspond to adiabatic ($\Omega=0$), resonant ($\Omega=h$), and fast ($\Omega=5h$) driving cases. We use $h=0.1$
meV, $T=0.05h$, $\varGamma^{a}_{\nu}=0.1$, and $V=5h$. The values of $v\left(t\right)$
and $i\left(t\right)-i_{0}\left(t\right)$ are normalized to $v$ and $g_{e}v$, respectively.
The inset shows the amplitude of the second harmonic of the current
$i^{\left(2\right)}$, normalized by $v^{2}$, as the function of
the driving frequency for the same $h$, $T$, $\varGamma^{a}_{\nu}$, and $V$. Since $i^{\left(2\right)}$
scales as $v^{2}$, curves for different amplitudes of the driving $v$ are indistinguishable.}
\label{fgr:LissajousAndBloch} 
\end{figure}

We now extend our analysis to the full dynamics of the current within
a driving period for the case (c) of Fig. \ref{fgr:Current} that
best models the conditions reported in Ref. \cite{Baumann2015}.
We study current response for driving frequencies $\Omega\ll h$ (adiabatic case) and
$\Omega\gg h$ (fast case) and compare it with the resonant case, $\Omega=h$.
For both adiabatic and fast driving we find that
\begin{align}
i\left(t\right)\simeq i_{0}\left(t\right)=\frac{dI}{dV}\left(V\right)\times v\,\sin\Omega t,
\end{align}
i.e., the ac response is purely ohmic.
The Lissajous curves depicting the nonohmic part of the response $i\left(t\right)-i_{0}\left(t\right)$ versus $v\left(t\right)=v\,\sin\Omega t$
are shown in Fig. \ref{fgr:LissajousAndBloch}(a) for different amplitudes and
frequencies of the driving. 

When the driving is adiabatic, i.e., $\Omega\ll h$, the instantaneous
current $i\left(t\right)$ is solely determined by the instantaneous
voltage at time $t$, in which case the Lissajous curve shows no hysteresis
and can be determined from the dc curve. On the other hand, when the
driving frequency is large, i.e., $\Omega\gg h$, the internal state
of the system has no time to adapt (see below). The magnetic moment
thus experiences vanishing time-averaged torque. The resulting conductivity,
that is determined by the state of the magnetic moment, also does
not depend on the time resulting in purely Ohmic response $i\left(t\right)\propto v\left(t\right)$.
At resonance $\Omega=h$, the $i-v$ characteristic
exhibits a hysteresis loop indicated by the non-vanishing area inside
the Lissajous curve. This shows that the non-linear processes
responsible for the generation of $i^{\left(0\right)}$ also induce higher
harmonics whose amplitudes $i^{\left(m\right)}$ increase at resonance.
The amplitude of the second harmonic $m=2$, shown in the inset of
Fig. \ref{fgr:LissajousAndBloch}(a), is comparable to $i^{\left(0\right)}$
and also scales as $v^{2}$.

To better understand the phenomena reported above, we investigate the
dynamics of the magnetic moment for three different regimes considered
above. Figure \ref{fgr:LissajousAndBloch} shows the orbit followed by the Bloch vector $\bs s=\left\{ \av{S_{x}},\av{S_{y}},\av{S_{z}}\right\} $ over a period of the drive. For adiabatic driving, $\Omega\ll h$,
shown in Fig. \ref{fgr:LissajousAndBloch}(b), the spin has
time to adapt to the applied voltage and its trajectory can be obtained by the static master equation.
The magnetization points in the $z$ direction and oscillates around the static (i.e., $v=0$) value with an amplitude that is proportional to $v$.
In the regime of fast driving,
$\Omega\gg h$, shown in Fig. \ref{fgr:LissajousAndBloch}(d), the magnetization
remains static and independent of $v$, acquiring the value obtained
in the static case for $v=0$. This can be simply explained by the
fact that, for the time scales experienced by the spin dynamics, the
alternating voltage averages out to zero. The resonant case, $\Omega=h$,
is shown in Fig. \ref{fgr:LissajousAndBloch}(c). The trajectories form circular
orbits almost parallel to the $xy$ plane, centered at the static value, and with radii proportional to $v$.
This shows that, at resonance, the perpendicular polarized current
exerts a spin-transfer torque that is able to excite the magnetic
moment of the atom. This process requires quantum coherence, as
it involves the elements of the reduced density matrix of the
magnetic moment that are off-diagonal with respect to the Hamiltonian.  

\begin{figure}[tb]
\includegraphics[width=\columnwidth]{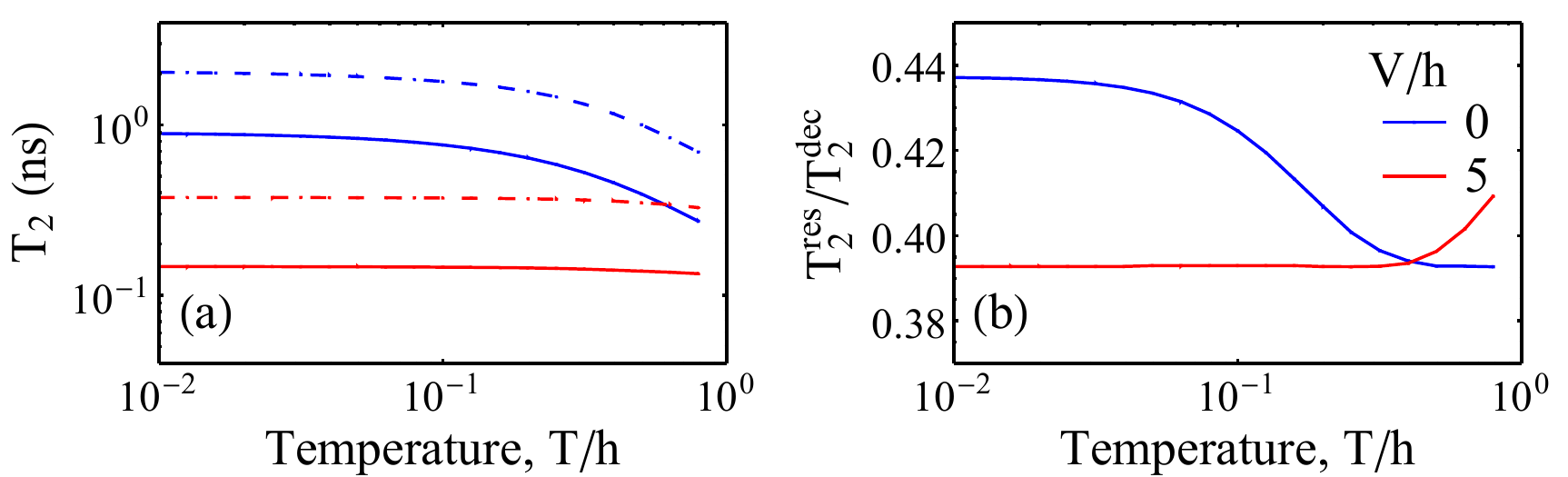}\caption{The dependence on the temperature of (a) the decoherence
times $T_{2}^{\t{res}}$ (solid curves) and $T_{2}^{\t{dec}}$ (dashed curves) estimated from the width of
the resonance and from the decay time of the coherence between the
spin states, correspondingly, (b) their ratio $T_{2}^{\t{res}}/T_{2}^{\t{dec}}$. The timescales $T_{2}^{\t{res}}$ are obtained
from fitting $i^{\left(0\right)}$ resonance curves for $h=0.1$ meV,
$v=0.1h$, $\varGamma^{a}_{\nu}=0.1$, and different values of $V$.}

\label{fgr:T2ratio} 
\end{figure}

Finally, we address another point raised by Ref. \cite{Baumann2015}
concerning the quantum coherence of a spin state. In Ref. \cite{Baumann2015},
the measurements of $i^{\left(0\right)}$ were used to indirectly evaluate
the decoherence time $T_{2}$ by determining the width of the resonance.
Such timescale can now be compared with the standard interpretation
of $T_{2}$ as the decay time of the coherences of the spin state \cite{Delgado2017}.
The definition of $T_{2}$ employed in the following has been established
in Ref. \cite{Shakirov2017}, where some of the subtleties of defining
a decoherence timescale in the presence of a spin-polarized environment
were addressed. This quantity, dubbed $T_{2}^{\t{dec}}$ in the following,
is determined by the fastest decay rate of information in a system
perturbed away from the non-equilibrium steady state that is established
in the presence of a static bias $V$. 

The analog of $T_{2}$ as measured in Refs. \cite{Baumann2015, Yang2017}, that
we denote $T_{2}^{\t{res}}$, is obtained from the width of the resonance
curves in Figs. \ref{fgr:Current}(a) and \ref{fgr:Current}(c), assuming that near
the resonance $i^{\left(0\right)}\left(\Omega\right)\propto e^{-b\left(\Omega-h\right)^{2}}$.
The half-height width is then computed as $T_{2}^{\t{res}}=\frac{1}{2}\sqrt{\frac{b}{\ln2}}$.
Figure \ref{fgr:T2ratio} depicts $T_{2}^{\t{res}}$, $T_{2}^{\t{dec}}$, and their ratio
as a function of the temperature for two values of $V$. The fact
that this ratio is constant at low temperature, and depends mildly
on the $V$, shows that $T_{2}^{\t{res}}$ can indeed be used to estimate
$T_{2}^{\t{dec}}$. However, both the variation observed at high temperatures
and the dependence on $V$ can yield a difference of the order of
$10\%$ in the ratio $T_{2}^{\t{res}}/T_{2}^{\t{dec}}$. This suggests
some caution to the assign a direct physical meaning to $T_{2}^{\t{res}}$.

The assumption of a spin-$1/2$ atom directly applies to Ref. \cite{Yang2017}. 
Nonetheless, our theoretical treatment and the effects it predicts generalize to higher magnetic moments, engineered spin structures, and magnetic molecules, providing  
the jump operators in Eq. (\ref{eq:LambdaOperators}) directly couple the resonant energy states. 
To avoid this limitation, a higher order expansion in the system-bath coupling has to be considered.

\section{Conclusion \label{sec:Conclusion}}

In this paper, we prove that spin-transfer torque may induce an electronic
paramagnetic resonance in single atomic spins. This mechanism does
not appeal to any mechanical or orbital degrees of freedom and only
involves the dissipative interaction of the polarized current with the atomic
spin. It is therefore independent of the environment of the atom which
may explain the ubiquity of the effect reported in Ref. \cite{Baumann2015}.
The current and spin dynamics induced by an ac voltage drive can be
well captured by a time-dependent master equation that generalizes
a previous framework \cite{Shakirov2016a} based on a Redfield-like
set of approximations. Our approach shows that the quantum coherence
of the atomic spin is crucial to capture paramagnetic resonance effects
resulting from the generation of second harmonics of the driving signal.
This nonlinear process is enhanced near the resonance condition and
depends on the square of the driving amplitude, which is compatible with Ref. \cite{Willke2018}. The  effect is based on a current-induced spin-torque and does not assume any effective time dependence of the local moment Hamiltonian. Moreover, the resonance is not observed for an unpolarized current which generates no spin-transfer torque.

We showed that the resonance width can reliably be used to estimate the decoherence timescale $T_{2}$ once mild temperature and voltage dependencies are accounted
for.

\begin{acknowledgments}
P.R. acknowledges support by FCT through the Investigador FCT Contract
No. IF/00347/2014 and Grant No. UID/CTM/04540/2013. We thankfully acknowledge
the computer resources, technical expertise and assistance provided
by CENTRA/IST. Computations were performed at the cluster Baltasar-Sete-S\'ois
and supported by the H2020 ERC Consolidator Grant, \textquotedbl Matter
and strong field gravity: New frontiers in Einstein's theory\textquotedbl{}
grant agreement No. MaGRaTh-646597.
The contribution of A.S. was funded by RFBR Grant No. 16-42-01057.
\end{acknowledgments}

\appendix

\section{Redfield equation}
\label{sec:Appendix}

In this Appendix, we derive the Redfield equation that was used
in the paper to calculate dynamics of the atom driven by the current.
The driving is included into the Hamiltonian of the environment as
a time-dependent shift to the energies of the electronic states
\begin{align}
H_{E}\left(t\right) & =\sum_{\nu ks}\left[\varepsilon_{\nu ks}-\delta_{\nu}\left(t\right)\right]c_{\nu ks}^{\dagger}c_{\nu ks}
\end{align}
where $\nu=s,t$ labels the leads, indices $k$ and $s$ enumerate
momentum and spin of the electrons, and $\Delta_{\nu}\left(t\right)=\Delta_{\nu}\sin\left(\Omega t\right)$.
We start by representing the coupling Hamiltonian, see Eq. (\ref{eq:CouplingHamiltonian}) of
the paper, as $H_{C}=\sum_{a}S_{a}\otimes E_{a}$ with
\begin{align}
E_{a} & =\sum_{\nu\nu'}\sqrt{\frac{J_{\nu}^{a}J_{\nu'}^{a}}{\mathcal{N}_{\nu}\mathcal{N}_{\nu'}}}\sum_{kk'ss'}c_{\nu ks}^{\dagger}\frac{\sigma_{ss'}^{a}}{2}c_{\nu'k's'},\label{eq:EnvironmentOperators}
\end{align}
which allows us to write the Redfield equation in the well-known form
\begin{align}
\partial_{t}\rho & =-i\left[H_{S}+\Delta H_{S},\rho\right]\\
& +\sum_{aa'}\left(\left[\varLambda_{aa'}\left(t\right)\rho,S_{a}\right]+\textrm{H.c.}\right).\nonumber
\end{align}
Here the Hamiltonian shift is given by
\begin{align}
\Delta H_{S} & =\sum_{a}\av{E_{a}}S_{a},
\end{align}
and the operators
\begin{align}
\varLambda_{aa'}\left(t\right) & =\int_{0}^{\infty}dt'e^{-iH_{S}t'}S_{a'}e^{iH_{S}t'}C_{aa'}\left(t,t'\right)\label{eq:LambdaOperators}
\end{align}
are expressed through the correlation functions of the environment
\begin{align}
C_{aa'}\left(t,t'\right) & =\tr_{E}\left[\left(E_{a}-\av{E_{a}}\right)e^{-i\int_{t-t'}^{t}H_{E}\left(\tau\right)d\tau}\label{eq:CorrelationFunctions}\right.\\
& \left.\times\left(E_{a'}-\av{E_{a'}}\right)e^{i\int_{t-t'}^{t}H_{E}\left(\tau\right)d\tau}\rho_{E}\right].\nonumber
\end{align}
Evaluating the Hamiltonian shift results in
\begin{align}
\Delta H_{S} & =\sum_{\nu a}\frac{J_{\nu}^{a}}{\mathcal{N}_{\nu}}\sum_{ks}\frac{\sigma_{ss}^{a}}{2}n_{F}\left(\varepsilon_{\nu ks}-\mu_{\nu}\right)S_{a}.
\end{align}
We employ spin-polarized DOS of the leads,
\begin{align}
\varrho_{\nu s}\left(\varepsilon\right) & =\frac{1}{\mathcal{N}_{\nu}}\sum_{k}\delta\left(\varepsilon-\varepsilon_{\nu ks}\right)\\
& =\frac{1}{2W}\left(1+p_{\nu}s\right)\Theta\left(|\varepsilon-\mu_{\nu}|-W\right),\nonumber
\end{align}
to replace sums over momenta by integrals over energy in this expression
and obtain
\begin{align}
\Delta H_{S} & =\frac{1}{2}\sum_{\nu}\int d\varepsilon\left[\left(J_{\nu}^{0}+J_{\nu}S_{z}\right)\varrho_{\nu\uparrow}\left(\varepsilon\right)\right.\\
& \left.+\left(J_{\nu}^{0}-J_{\nu}S_{z}\right)\varrho_{\nu\downarrow}\left(\varepsilon\right)\right]n_{F}\left(\varepsilon-\mu_{\nu}\right).\nonumber
\end{align}
In the large bandwidth limit, with constant terms discarded, one gets
\begin{align}
\Delta H_{S} & =\frac{1}{2}\sum_{\nu}J_{\nu}p_{\nu}S_{z},
\end{align}
or $\Delta H_{S}=\frac{1}{2}J_{t}pS_{z}$ for the case when only tip
is polarized. Because of driving, the correlation functions depend
on both time arguments rather than their difference. Substituting
Eq. (\ref{eq:EnvironmentOperators}) into Eq. (\ref{eq:CorrelationFunctions})
results in
\begin{align}
C_{aa'}\left(t,t'\right) & =\sum_{\nu\nu'}\frac{\sqrt{J_{\nu}^{a}J_{\nu'}^{a}J_{\nu}^{a'}J_{\nu'}^{a'}}}{\mathcal{N}_{\nu}\mathcal{N}_{\nu'}}\sum_{kk'ss'}\frac{\sigma_{ss'}^{a}\sigma_{s's}^{a'}}{4}\\
 & \times\left\langle c_{\nu ks}^{\dagger}c_{\nu'k's'}e^{-i\int_{t-t'}^{t}H_{E}\left(\tau\right)d\tau}\right.\nonumber\\
 & \left.\times c_{\nu'k's'}^{\dagger}c_{\nu ks}e^{i\int_{t-t'}^{t}H_{E}\left(\tau\right)d\tau}\right\rangle.\nonumber 
\end{align}
Introducing DOS of the leads, we replace sums over momenta by integrals
over energy in this expression and obtain
\begin{align}
C_{aa'}\left(t,t'\right) & =\sum_{\nu\nu'}\sqrt{J_{\nu}^{a}J_{\nu'}^{a}J_{\nu}^{a'}J_{\nu'}^{a'}}\sum_{ss'}\frac{\sigma_{ss'}^{a}\sigma_{s's}^{a'}}{4}\\
& \times\int\int d\varepsilon d\varepsilon'\varrho_{\nu s}\left(\varepsilon\right)\varrho_{\nu's'}\left(\varepsilon'\right)\nonumber\\
& \times e^{i\int_{t-t'}^{t}\left[\varepsilon-v_{\nu}\left(\tau\right)\right]d\tau}e^{-i\int_{t-t'}^{t}\left[\varepsilon'-v_{\nu'}\left(\tau\right)\right]d\tau}\nonumber\\
& \times n_{F}\left(\varepsilon-\mu_{\nu}\right)\left[1-n_{F}\left(\varepsilon-\mu_{\nu'}\right)\right].\nonumber 
\end{align}
We then employ rectangular DOS, introduce dimensionless coupling parameters
$\varGamma_{\nu}^{a}=\pi J_{\nu}^{a}/\left(2W\right)$, and use the
relation $1-n_{F}\left(\varepsilon\right)=n_{F}\left(-\varepsilon\right)$
to rewrite the last expression as
\begin{align}
C_{aa'}\left(t,t'\right) & =\sum_{\nu\nu'}\frac{1}{4\pi}\sqrt{\varGamma_{\nu}^{a}\varGamma_{\nu'}^{a}\varGamma_{\nu}^{a'}\varGamma_{\nu'}^{a'}}\label{eq:CorrelationFunctionsFinal}\\
& \times\tr\left[\left(1+p_{\nu}\sigma^{z}\right)\sigma^{a}\left(1+p_{\nu'}\sigma^{z}\right)\sigma^{a'}\right]\nonumber\\
& \times e^{i\left(\mu_{\nu}-\mu_{\nu'}\right)t'-i\int_{t-t'}^{t}\left[v_{\nu}\left(\tau\right)-v_{\nu'}\left(\tau\right)\right]d\tau}\nonumber\\
& \times\frac{1}{\pi}\left[\int_{-W}^{W}d\varepsilon e^{i\varepsilon t'}n_{F}\left(\varepsilon\right)\right]^{2}.\nonumber 
\end{align}
Let us introduce the definition
\begin{align}
u_{aa'}^{\nu\nu'} & =\frac{1}{4\pi}\sqrt{\varGamma_{\nu}^{a}\varGamma_{\nu'}^{a}\varGamma_{\nu}^{a'}\varGamma_{\nu'}^{a'}}\\
& \times\tr\left[\left(1+p_{\nu}\sigma^{z}\right)\sigma^{a}\left(1+p_{\nu'}\sigma^{z}\right)\sigma^{a'}\right]\nonumber
\end{align}
and substitute Eq. (\ref{eq:CorrelationFunctionsFinal}) into Eq. (\ref{eq:LambdaOperators})
using the spectral decomposition $I=\sum_{\alpha}\ket{\alpha}\bra{\alpha}$.
We get
\begin{align}
\varLambda_{aa'}\left(t\right) & =\sum_{\alpha\alpha'}\ket{\alpha}\bra{\alpha}S_{a}\ket{\alpha'}\bra{\alpha'}\\
& \times\sum_{\nu\nu'}u_{aa'}^{\nu\nu'}\int_{0}^{\infty}dt'e^{-i\left(\omega_{\alpha}-\omega_{\alpha'}-\mu_{\nu}+\mu_{\nu'}\right)t'}\nonumber\\
& \times e^{-i\int_{t-t'}^{t}\left[v_{\nu}\left(\tau\right)-v_{\nu'}\left(\tau\right)\right]d\tau}\nonumber\\
& \times\frac{1}{\pi}\left[\int_{-W}^{W}d\varepsilon e^{i\varepsilon t'}n_{F}\left(\varepsilon\right)\right]^{2}\nonumber 
\end{align}
which coincides with Eq. (\ref{eq:LambdaOperators}) of the paper, where
\begin{align}
\kappa_{\nu\nu'}^{t}\left(\omega\right) & =\int_{0}^{\infty}dt'e^{-i\left(\omega-\mu_{\nu}+\mu_{\nu'}\right)t'}\label{eq:KappaTimeDependent}\\
& \times e^{-i\int_{t-t'}^{t}\left[v_{\nu}\left(\tau\right)-v_{\nu'}\left(\tau\right)\right]d\tau}\nonumber\\
& \times\frac{1}{\pi}\left[\int_{-W}^{W}d\varepsilon e^{i\varepsilon t'}n_{F}\left(\varepsilon\right)\right]^{2}.\nonumber
\end{align}
For sinusoidal periodic driving $v_{\nu}\left(t\right)=v_{\nu}\sin\left(\Omega t\right)$,
we may decompose
\begin{align}
& e^{-i\int_{t-t'}^{t}\left[v_{\nu}\left(\tau\right)-v_{\nu'}\left(\tau\right)\right]d\tau}\label{eq:BesselDecomposition}\\
& =e^{i\frac{v_{\nu}-v_{\nu'}}{\Omega}\left[\cos\left(\Omega t\right)-\cos\left(\Omega\left(t-t'\right)\right)\right]}\nonumber\\
& =e^{i\frac{v_{\nu}-v_{\nu'}}{\Omega}\cos\left(\Omega t\right)}\sum_{m=-\infty}^{+\infty}i^{m}J_{m}\left(-\frac{v_{\nu}-v_{\nu'}}{\Omega}\right)e^{im\Omega\left(t-t'\right)},\nonumber 
\end{align}
where we used the identity $e^{iz\cos\varphi}=\sum_{m=-\infty}^{+\infty}i^{m}J_{m}\left(z\right)e^{im\varphi}$.
Substituting this into Eq. (\ref{eq:KappaTimeDependent}) gives Eq. (\ref{eq:KappaFunctionTime}) of the paper with
\begin{align}
\kappa & \left(\omega\right)=\frac{1}{\pi}\int_{0}^{\infty}dt\,e^{-i\omega t}\left[\int_{-W}^{W}d\varepsilon e^{i\varepsilon t}n_{F}\left(\varepsilon\right)\right]^{2}.
\end{align}
The evaluation of this integral in the large bandwidth limit $W\to\infty$ results in
\begin{align}
\kappa\left(\omega\right) & =-\frac{i}{\pi}\left(bW+\omega\ln\frac{\left|\omega\right|}{cW}\right)\\
& +\frac{g\left(\beta\omega\right)+i\,f\left(\beta\omega\right)}{\beta}+O\left(\frac{\left|\omega\right|}{W}\right),\nonumber
\end{align}
where $b=2\ln2$ and $c=e/2$. We note that DOS with different from
rectangular shapes give $\kappa\left(\omega\right)$ of the same form
but with other values of $b$ and $c$, e.g., for $\varrho_{\nu s}\left(\varepsilon\right)=\frac{1}{2W}\left(1+p_{\nu}s\right)\exp\left(-\left|\varepsilon-\mu_{\nu}\right|/W\right)$
one gets $b=1$ and $c=e^{-\gamma}$. The term proportional to the
bandwidth gives no contribution to the equation and we thus exclude
it from $\kappa\left(\omega\right)$, as well as $O\left(\left|\omega\right|/W\right)$
term, and arrive at Eq. (\ref{eq:KappaFunction}) of the paper.

\bibliographystyle{apsrev4-1}
\bibliography{DrivenBib}

\end{document}